\documentclass [12pt,a4paper]{article}
\usepackage{amssymb, theorem}
\newtheorem{thm}{Theorem}
\newtheorem{cor}{Corollary}

\newtheorem{lemma}{Lemma}
\theorembodyfont{\rm}
\newtheorem{pl}{Example}

\def\proof{{\it Proof: }}
\def\Cov{\mathrm{Cov}}

\def\QCov{\mathrm{qCov}}

\def\qed{\nobreak\hfill $\square$}
\def\<{\langle}
\def\>{\rangle}
\def\aa{\alpha}

\def\fel{{\textstyle{1 \over 2}}}

\def\bL{{\bf L}}

\def\bM{{\bf M}}
\def\iM{{\cal M}}

\def\ot{\otimes}
\def\osum{\oplus}

\def\bbbr{{\mathbb R}}

\def\Diag{\mbox{Diag}\,}

\def\Tr{\mathrm{Tr}\,}

\def\L{{\mathbb L}}
\def\J{{\mathbb J}}

\def\bL{{\mathbb L}}
\def\bR{{ \mathbb R}}

\def\fel{\textstyle{\frac{1}{2}}}

\begin{document}
\ \vskip 1cm 
\centerline{\LARGE {\bf From quasi-entropy}}
\bigskip
\bigskip
\bigskip
\centerline{D\'enes Petz\footnote{E-mail: petz@math.bme.hu.
Partially supported by the Hungarian Research Grant OTKA  T068258 and
the Mittag-Leffler Institute in Stockholm.}}
\centerline{Alfr\'ed R\'enyi Institute of Mathematics}
\centerline{H-1364 Budapest, POB 127, Hungary}
\bigskip
\bigskip
\begin{abstract}
The subject is the overview of the use of quasi-entropy in finite dimensional
spaces. Matrix monotone functions and relative modular operators are used. The origin
is the relative entropy and the $f$-divergence, monotone metrics, covariance and the 
$\chi^2$ divergence are the most important particular cases. The extension of the 
monotone metric to two variables is a new concept.

\medskip\noindent
{\bf Key words and phrases:}
$f$-divergence, quasi-entropy, von Neumann entropy, relative entropy,
monotonicity property, Fisher information, $\chi^2$-divergence.
\end{abstract}

Quasi-entropy was introduced by Petz in 1985 as the quantum generalization
of Csisz\'ar's $f$-divergence in the setting of matrices or von Neumann
algebras. The important special case was the relative entropy of Umegaki
and Araki. In this paper the applications are overviewed in the finite 
dimensional setting. Quasi-entropy has some similarity to the monotone
metrics, in both cases the modular operator is included, but there is an 
essential difference: In the  quasi-entropy two density matrices are
included and for the monotone metric on foot-point density matrices. In
this paper two density matrices are introduced in the monotone metric
style.


\section{Quasi-entropy}

Let $\iM$ denote the algebra of $n \times n$ matrices with complex entries.
For positive definite matrices $\rho_1, \rho_2\in \iM$, for $A \in \iM$ and 
a function $f:\bbbr^+ \to \bbbr$, the {\it quasi-entropy} is defined as
\begin{eqnarray}\label{E:quasi}
S^A_f (\rho_1\|\rho_2)&:=& \< A \rho_2^{1/2}, f(\Delta(\rho_1/ \rho_2))
(A\rho_2^{1/2})\> \cr
&=&\Tr  \rho_2^{1/2} (A^*f(\Delta(\rho_1/ \rho_2))A\rho_2^{1/2}),
\end{eqnarray}
where $\<B,C\>:=\Tr B^*C$ is the so-called {\it Hilbert-Schmidt inner product} 
and $\Delta(\rho_1/ \rho_2):\iM \to \iM$ is a linear mapping acting 
on matrices:
$$
\Delta(\rho_1/ \rho_2)B=\rho_1 B\rho_2^{-1}.
$$
This concept was introduced by Petz in 1985, see \cite{PD26, PD32}, or Chapter 7 
in \cite{OP}. (The relative modular operator  $\Delta(\rho_1/ \rho_2)$ was born in
the context of von Neumann algebras and the paper of Araki \cite{Araki} had a
big influence even in the matrix case.)
The quasi-entropy is the quantum generalization of the $f$-divergence of 
Csisz\'ar used in classical information theory (and statistics) \cite{Csi, LV}.
Therefore the quantum $f$-divergence could be another terminology as in \cite{HMP}.

The definition of quasi-entropy can be formulated with mean. For a function $f$
the corresponding mean is defined as $m_h(x,y)=f(x/y)y$ for positive numbers, or
for commuting positive definite matrices. The linear mappings
$$
L_{\rho_1}X=\rho_1 X \quad \mbox{and} \quad R_{\rho_2}X=X\rho_2
$$
are positive and commuting. The mean $m_f$ makes sense and
\begin{eqnarray}\label{E:qmean}
S^A_f (\rho_1\|\rho_2) = \< A,m_f (\L_{\rho_1}, R_{\rho_2})A\>.
\end{eqnarray}

Let $\alpha :\iM_0 \to \iM$ be a mapping between two matrix algebras. The dual
$\alpha^*: \iM \to \iM_0$ with respect to the Hilbert-Schmidt inner product
is positive if and only if $\alpha$ is positive. Moreover, $\alpha$ is
unital if and only if $\alpha^*$ is trace preserving. $\alpha: \iM_0 \to \iM$
is called a {\it Schwarz mapping} if 
\begin{equation}\label{E:Sch}
\alpha(B^*B)\ge \alpha(B^*)\alpha(B)
\end{equation}
for every $B \in \iM_0$. 

The quasi-entropies are monotone and jointly convex \cite{OP, PD32}.

\begin{thm}\label{T:quasimon}
Assume that $f:\bbbr^+ \to \bbbr$ is an operator monotone function
with $f(0)\ge 0$ and $\alpha:\iM_0 \to \iM$ is a unital Schwarz mapping. 
Then
\begin{equation}\label{E:quasimonB}
S^A_f (\alpha^*(\rho_1) \| \alpha^*(\rho_2)) \ge
S^{\alpha(A)}_f (\rho_1 \| \rho_2)
\end{equation}
holds for $A \in \iM_0$ and for invertible density matrices $\rho_1$ and 
$\rho_2$ from the matrix algebra $\iM$.
\end{thm}

\proof
The proof is based on inequalities for operator monotone and
operator concave functions. First note that
$$
S^A_{f+c} (\alpha^*(\rho_1)\| \alpha^*(\rho_2))=
S^A_f (\alpha^*(\rho_1)\| \alpha^*(\rho_2))+c\,\Tr \rho_1\alpha(A^*A))
$$
and 
$$
S^{\alpha(A)}_{f+c} (\rho_1\| \rho_2)=S^{\alpha(A)}_f (\rho_1\|\rho_2)+
c\,\Tr \rho_1 (\alpha(A)^*\alpha(A))
$$
for a positive constant $c$. Due to the Schwarz inequality (\ref{E:Sch}),
we may assume that $f(0)=0$.

Let $\Delta:=\Delta(\rho_1/ \rho_2)$ and $\Delta_0:=\Delta(\alpha^*(\rho_1)
/ \alpha^*(\rho_2))$. The operator
\begin{equation}
VX\alpha^*(\rho_2)^{1/2}=\alpha(X)\rho_2^{1/2} \qquad (X \in \iM_0)
\end{equation}
is a contraction:
\begin{eqnarray*}
\Vert \alpha(X)\rho_2^{1/2} \Vert^2 &=& \Tr \rho_2 (\alpha(X)^* 
\alpha(X)) \cr &\le&
\Tr \rho_2 (\alpha(X^*X)= \Tr \alpha^*(\rho_2) X^*X = 
\Vert X \alpha^*(\rho_2)^{1/2}\Vert^2
\end{eqnarray*}
since the Schwarz inequality is applicable to $\alpha$. A similar simple
computation gives that
\begin{equation}
V^*\Delta V \le \Delta_0\,.
\end{equation}

Since $f$ is operator monotone, we have $f(\Delta_0)\ge f(V^*\Delta V)$. 
Recall that $f$ is operator concave, therefore $f(V^*\Delta V) \ge 
V^*f(\Delta)V$ and we conclude
\begin{equation}
f(\Delta_0) \ge V^*f(\Delta)V\,.
\end{equation}
Application to the vector $A \alpha^*(\rho_2)^{1/2}$ gives the statement. \qed

It is remarkable that for a multiplicative $\alpha$ we do not need the 
condition $f(0)\ge 0$. Moreover, $V^*\Delta V = \Delta_0$ and we do not 
need the matrix monotonicity of the function $f$. In this case the only
condition is the matrix concavity, analogously to Theorem \ref{T:quasimon}.
If we apply the monotonicity (\ref{E:quasimonB}) to the embedding
$\alpha(X)=X \osum X$ of $\iM$ into $\iM \osum \iM$ and to the
densities $\rho_1=\lambda E_1\osum(1-\lambda)F_1$, $\rho_2=\lambda E_2
\osum (1-\lambda)F_2$, then we obtain the joint concavity of the 
quasi-entropy:

\begin{thm}\label{T:quasiconv}
If $f:\bbbr^+ \to \bbbr$ is an operator convex, then $S^A_f (\rho_1\| \rho_2)$
is jointly convex in the variables $\rho_1$ and $\rho_2$.
\end{thm}

If we consider the quasi-entropy in the terminology of means, then we can have another
proof. The joint convexity of the mean is the inequality
$$
f(L_{(A_1+A_2)/2}R_{(B_1+B_2)/2}^{-1})R_{(B_1+B_2)/2} \le \fel f(L_{A_1}R_{B_1}^{-1})R_{B_1}+
\fel f(L_{A_2}R_{B_2}^{-1})R_{B_2}
$$
which can be simplified as
\begin{eqnarray*}
& & f(L_{A_1+A_2}R_{B_1+B_2}^{-1}) \cr & & \qquad \cr & & \qquad
\le R_{B_1+B_2}^{-1/2} R_{B_1}^{1/2} f(L_{A_1}R_{B_1}^{-1})R_{B_1}^{1/2} 
R_{B_1+B_2}^{-1/2}+ R_{B_1+B_2}^{-1/2}R_{B_2}^{1/2} f(L_{A_2}R_{B_2}^{-1})R_{B_2}^{1/2} R_{B_1+B_2}^{-1/2}
\cr & & \qquad \cr & & \qquad
\le C f(L_{A_1}R_{B_1}^{-1})C^* + D f(L_{A_2}R_{B_2}^{-1})D^*.
\end{eqnarray*}
Here $CC^*+DD^*=I$ and
$$
C (L_{A_1}R_{B_1}^{-1})  C^*+D (L_{A_2}R_{B_2}^{-1})  D^*=L_{A_1+A_2}R_{B_1+B_2}^{-1}.
$$
So the joint convexity of the quasi-entropy has the form
$$
f(CXC^*+ DYD^*) \le C f(X) C^*+ Df(Y)D^*
$$
which is true for an operator convex function $f$ \cite{HP, pd2}.

If $f$ is operator monotone function, then it is operator concave and we have joint
concavity in the previous theorem. The book \cite{pd2} contains information about 
operator monotone functions. The standard useful properties are integral representations.
The L\"owner theorem is
$$
f(x)=f(0)+\beta x+ \int_0^\infty\frac{\lambda x}{\lambda +x}\,d\mu(\lambda)\,. 
$$

An operator monotone function $f:\bbbr^+ \to \bbbr^+$ will be called 
{\it standard} if $xf(x^{-1})=f(x)$ and $f(1)=1$. A standard function $f$ 
admits a canonical representation

\begin{equation}\label{E:canonicalf}
f(t)=\frac{1+t}{2}\exp
\int_0^1(1-t)^2\frac{\lambda^2-1}{(\lambda+t)(1+\lambda t)(\lambda+1)^2
}h(\lambda)\,d\lambda,
\end{equation}
where $ h:[0,1]\to[0,1] $ is a measurable function \cite{H1}.  


\section{Applications}

The concept of quasi-entropy includes many important special cases.

\subsection{$f$-divergences}

If $\rho_2$ and $\rho_1$ are different and $A=I$, then we have a kind of relative 
entropy. For $f(x)= x\log x$ we have Umegaki's relative entropy 
$S(\rho_1\|\rho_2)=\Tr \rho_1 (\log \rho_1 - \log \rho_2)$. (If we want a
matrix monotone function, then we can take $f(x)=\log x$ and then we get
$S(\rho_2\|\rho_1)$.) Umegaki's relative entropy is the most important 
example, therefore the function $f$ will be chosen to be matrix convex.
This makes the probabilistic and non-commutative situation compatible
as one can see in the next argument.

Let $\rho_1$ and $\rho_2$ be density matrices in $\iM$. If in certain
basis they have diagonal $p=(p_1.p_2, \dots ,p_n)$ and $q=(q_1,q_2, 
\dots ,q_n)$, then the monotonicity theorem gives the inequality
\begin{equation}\label{E:sub}
D_f(p\|q) \le S_f(\rho_1\|\rho_2)
\end{equation}
for a matrix convex function $f$. If $\rho_1$ and $\rho_2$ commute, them we 
can take the common eigenbasis and  in (\ref{E:sub}) the equality appears. 
It is not trivial that otherwise the inequality is strict.

If $\rho_1$ and $\rho_2$ are different, then there is a choice for $p$ and $q$
such that they are different as well. Then
$$
0< D_f(p\|q) \le S_f(\rho_1\|\rho_2).
$$
Conversely, if $S_f(\rho_1\|\rho_2)=0$, then $p=q$ for every basis and this
implies $\rho_1=\rho_2$. For the relative entropy, a deeper result is known.
The {\it Pinsker-Csisz\'ar inequality} says that
\begin{equation}\label{E:PCs}
\|p-q\|_1^2 \le 2 D(p\|q).
\end{equation}
This extends to the quantum case as
\begin{equation}\label{E:HO}
\|\rho_1-\rho_2\|_1^2 \le 2 S(\rho_1\|\rho_2),
\end{equation}
see \cite{HOT}, or \cite[Chap. 3]{pd2}.

\begin{pl}\label{P:1}
The $f$-divergence with $f(x)=x \log x$ is the relative entropy. It is rather popular
the modification of the logarithm as
$$
\log_\beta x = \frac{x^\beta -1}{\beta} \qquad (\beta \in (0,1))
$$
and the limit $\beta \to 0$ is the $\log$. If we take $f_\beta(x)=x \log_\beta x$, then
$$
S_\beta (\rho_1\|\rho_2)=\frac{\Tr \rho_1^{1+\beta} \rho_2^{-\beta} -1}{\beta}.
$$ 
Since $f_\beta$ is operator convex, this is a good generalized entropy. It appeared in
the paper \cite{RS}, see also \cite[Chap. 3]{OP}, there $\gamma$ is written instead 
of $\beta$ and
$$
S(\rho_1\|\rho_2) \le S_\beta (\rho_1\|\rho_2) \qquad (\beta \in (0,1))
$$
is proven.

The {\it relative entropies of degree} $\alpha$  
$$
S_\alpha(\rho_2\|\rho_1):={1 \over \aa(1-\aa)}
\Tr (I-\rho_1^{\aa}\rho_2^{-\aa})\rho_2.
$$
are essentially the same. \qed
\end{pl}

The $f$-divergence is contained in details in the recent papers \cite{PD138, HMP}.

\subsection{WYD information}

In the paper \cite{JR} the functions
$$
g_p(x) =  \cases{ \frac{1}{p(1-p)} (x - x^p)  & if $p  \neq 1$,
\cr \cr   x \log x  &  if $p = 1$ }
$$
are used, this is a reparametrization of Example \ref{P:1}. (Note that $g_p$ is 
well-defined for $x > 0$ and $p \neq 0$.) The considered case is $p \in [1/2,2]$, then 
$g_p$ is operator concave. 

For strictly positive $A$ and $B$,  Jen\v{c}ov\'{a} and Ruskai define 
$$
J_p(K,A,B) = \Tr \sqrt{B} K^*  \, g_p\big(L_A R_B^{-1}\big)  (K  \sqrt{B})
$$
which is the particular case of the quasi-entropy $S^K_f (A\|B)$ with $f=g_p$.

The joint concavity of $J_p(K,A,B)$ is stated in Theorem 2 in \cite{JR}
and this is a particular case of Theorem \ref{T:quasiconv} above.
For $K = K^*$, we have
$$
J_p(K,A,A) = -  \frac{1}{2p(1-p)} \Tr [K,A^p][K,A^{1-p}]
$$
which is the Wigner-Yanase-Dyson  information (up to a constant) and extends it to the range 
$(0,2]$.

\subsection{Monotone metrics}

Let $\iM_n$ be the set of positive definite density matrices in $\bM_n$. This is
a manifold and the set of tangent vectors is $\{A=A^* \in \bM_n\,:\, \Tr A=0\}$.
A Riemannian geometry is a set of real inner products $\gamma_D(A,B)$ on the 
tangent vectors \cite{Naga}. By monotone metrics we mean inner product for all matrix spaces
such that
\begin{equation}\label{E:Fmon}
\gamma_{\beta(D)}(\beta(A),\beta(A) )  \le \gamma_D(A,A)
\end{equation}
for every completely positive trace preserving mapping $\beta:\bM_n \to \bM_m$.

Define $\J_D^f :\bM_n \to \bM_n$ as
\begin{equation}\label{E:jede}
\J_D^f=f(\bL_D\bR_D^{-1})\bR_D= \bL_D \, m_f \, \bR_D \,,
\end{equation}
where $f: \bbbr^+ \to \bbbr^+$ and $m_f$ is the mean induced by the function 
$f$.

It was obtained in the paper \cite{PD2} that monotone
metrics with the property
\begin{equation}\label{E:Fish}
\gamma_D (A, A)=\Tr D^{-1}A^2 \quad \mbox{if} \quad AD=DA
\end{equation}
has the form
\begin{equation}\label{E:jede2}
\gamma_D(A,B)=\Tr A(\J_D^f)^{-1}(B)
\end{equation}
where $f$ is a standard matrix monotone function. These monotone metrics are
abstract Fisher informations, the condition (\ref{E:Fish}) tells that in the commutative
case the classical Fisher information is required. The popular case in physics corresponds 
to $f(x)=(1+x)/2$, this gives the SSA Fisher information.

Since
$$
\Tr A(\J_D^f)^{-1}(B)=\< (AD^{-1})D^{1/2}, \frac{1}{f}(\Delta(D/D))(AD^{-1})D^{1/2}\>,
$$
we have
$$
\gamma_D(A,A)=S^{AD^{-1}}_{1/f} (D\|D).
$$
So the monotone metric is a particular case of the quasi-entropy, but there is
another relation. The next example has been well-known.

\begin{pl}
The Boguliubov-Kubo-Mori Fisher information is induced by the function
$$
f(x)=\frac{x-1}{\log x}=\int_0^1 x^t\,dt.
$$ 
Then
$$
\J_D^f A= \int_0^1 (\bL_D\bR_D^{-1})^t\bR_D A\,dt=\int_0^1 D^t AD^{1-t}\, dt
$$
and computing the inverse we have
$$
\gamma_D^{BKM}(A,A)= \int_0^\infty \Tr (D+tI)^{-1} A (D+tI)^{-1}A\, dt.
$$
A characterization is in the paper \cite{G-S} and the relation with the relative 
entropy is
$$
\gamma_D^{BKM}(A,B)=\frac{\partial^2}{\partial t \partial s}
S(D+tA\|D+sB).
$$
\qed \end{pl}

Ruskai and Lesniewski discovered that all monotone Fisher informations
are obtained from an $f$-divergence by derivation \cite{L-R}:
$$
\gamma_D^f(A,B)=\frac{\partial^2}{\partial t \partial s}
S_F(D+tA\|D+sB)
$$
The relation of the function $F$  to the function $f$ in this formula is 
\begin{equation}\label{E:Ruskfor}
\frac{1}{f(t)}=\frac{F(t)+tF(t^{-1})}{(t-1)^2}.
\end{equation}

If $D$ runs on all positive definite matrices, conditions $\gamma_D(A,A) \in \bbbr$
for self-adjoint $A$ and (\ref{E:Fish}) are not required, but the monotonicity
(\ref{E:Fmon}) is assumed, then we have the generalized monotone metric characterized
by Kumagai \cite{Wataru}. They have the form
$$
K_\rho (A,B)=b(\Tr \rho)\Tr A^* \Tr B+ c \<A, (\J_\rho^f)^{-1} (B)\>,
$$
where $f:\bbbr^+ \to \bbbr^+$ is matrix monotone, $f(1)=1$, $b:\bbbr^+ \to \bbbr^+$
and $c>0$.

Let $\beta: \bM_n \ot \bM_2 \to \bM_m$ be defined as
$$
\left[\matrix{B_{11} & B_{12} \cr B_{21} & B_{22}}\right] \mapsto B_{11}+B_{22}.
$$
This is completely positive and trace-preserving, it is a so-called partial trace. For
$$
D=\left[\matrix{\lambda D_{1} & 0 \cr 0 & (1-\lambda)D_{2}}\right], \qquad
A=\left[\matrix{\lambda B  & 0 \cr 0 & (1-\lambda)B}\right]
$$
the inequality (\ref{E:Fmon}) gives
$$
\gamma_{\lambda D_1+(1-\lambda)D_2}(B,B)  \le \gamma_{\lambda D_1}(\lambda B,\lambda B)+
\gamma_{(1-\lambda) D_2}((1-\lambda) B,(1-\lambda) B).
$$
Since $\gamma_{tD} (tA, tB)=t\gamma_D (A, B)$, we obtained the convexity.

\begin{thm}\label{T:conv}
For a  standard matrix monotone function $f$ and for a self-adjoint matrix $A$
the monotone metric $\gamma_{D}^f (A, A)$ is a convex function of $D$.
\end{thm}

This convexity relation can be reformulated from formula (\ref{E:jede2}). We
have the convexity of the operator $(\J_D^f)^{-1}$ in the positive definite $D$.

\subsection{Generalized covariance}

If $\rho_2=\rho_1=\rho$ and $A, B \in \iM$ are arbitrary, then one can 
approach  to the {\it generalized covariance} \cite{PD22}.
\begin{equation}\label{E:qC}
\QCov^f_{\rho}(A,B):=\< A \rho ^{1/2}, f(\Delta(\rho / \rho ))(B\rho ^{1/2})\>
-(\Tr \rho A^*)(\Tr \rho B).
\end{equation}
is a generalized covariance. The first term is $\< A, \J_\rho^f B\>$ and the covariance 
has some similarity to the monotone metrics.

If $\rho, A$ and $B$ commute, then this 
becomes $f(1) \Tr \rho A^*B-(\Tr \rho A^*)(\Tr \rho B)$. This shows that
the normalization $f(1)=1$ is natural. The generalized covariance 
$\QCov^f_{\rho}(A,B)$ is a sesquilinear form and it is determined by 
$\QCov^f_{\rho}(A,A)$ when $\{ A\in \iM: \Tr \rho A=0\}$. Formally, 
this is a quasi-entropy and Theorem \ref{T:quasimon} applies if 
$f$ is matrix monotone. If we require the symmetry condition 
$\QCov^f_{\rho}(A,A)=\QCov^f_{\rho}(A^*,A^*)$, then $f$ should have
the symmetry $xf(x^{-1})=f(x)$.

Assume that $\Tr \rho A=\Tr \rho B=0$ and $\rho=\Diag(\lambda_1, \lambda_2,\dots,
\lambda_n)$. Then
\begin{equation}\label{E:qC2}
\QCov^f_{\rho}(A,B)=\sum_{ij} \lambda_i f (\lambda_j/\lambda_i)
A^*_{ij}B_{ij}.
\end{equation}

The usual {\it symmetrized covariance} corresponds to the function 
$f(t)=(t+1)/2$:
$$
\Cov_\rho (A,B):=
\frac{1}{2}\Tr (\rho (A^*B+BA^*))- (\Tr \rho A^*)(\Tr \rho B).
$$

The interpretation of the covariances is not at all clear. In the next
section they will be called {\it quadratic cost functions}. It turns out that
there is a one-to-one correspondence between quadratic cost functions
and Fisher informations.

\begin{thm}\label{T:conc}
For a  standard matrix monotone function $f$ the covariance $\QCov^f_{\rho}(A,A)$ is a 
concave function of $\rho$ for a self-adjoint $A$.
\end{thm}

\proof
The argument similar to the proof of Theorem \ref{T:conv}. Instead of the inequality
$\beta^* (\J_{\beta(D)}^f)^{-1}\beta  \le (\J_D^f)^{-1}$ we use the inequality
$\beta \J_{D}^f\beta^*  \le \J_{\beta(D)}^f$ (see Theorem 1.2 in \cite{PD143} or \cite{PD22}). 
This gives the concavity of $\< A, \J_\rho^f a\>$. The convexity of $(\Tr \rho A)^2$ is 
obvious.\qed


\subsection{$\chi^2$-divergence} 

The $\chi^2$-divergence 
$$
\chi^2(p,q)=\sum_i\frac{(p_i-q_i)^2}{q_i}=\sum_i\left(\frac{p_i}{q_i}-1\right)^2q_i
$$
was first introduced by Karl Pearson in 1900. 
Since
$$
\left(\sum_i |p_i -q_i|\right)^2 =
\left(\sum_i\left|\frac{p_i}{q_i}-1\right|q_i\right)^2 \le
\sum_i\left(\frac{p_i}{q_i}-1\right)^2q_i,
$$
we have
\begin{equation}\label{E:chinorm}
\|p-q\|_1^2 \le \chi^2(p,q).
\end{equation}
We also remark that the $\chi^2$-divergence is an $f$-divergence of Csisz\'ar with
$f(x)=(x-1)^2$ which is a (matrix) convex function. In the quantum case definition 
(\ref{E:quasi}) gives
$$
S_f(\rho,\sigma)=\Tr \rho^2 \sigma^{-1} - 1.
$$
Another quantum generalization was introduced very recently in \cite{TRus}:
$$
\chi^2_\alpha(\rho,\sigma) = \Tr \left(\rho-\sigma)\sigma^{-\alpha} 
(\rho-\sigma)\sigma^{\alpha-1} \right)=\Tr \rho \sigma^{-\alpha}\rho \sigma^{\alpha-1} -1
$$
where $\alpha \in [0,1]$. If $\rho$ and $\sigma$ commute, then this formula is 
independent of $\alpha$. In the general case the above $S_f(\rho,\sigma)$ comes
for $\alpha=0$.

More generally, they defined
$$
\chi^2_k(\rho,\sigma):=\left \langle \rho-\sigma,\Omega^k_\sigma(\rho-\sigma) 
\right \rangle, 
$$
where $\Omega^k_\sigma = R^{-1}_{\sigma}k(\Delta (\sigma/\sigma))$ and $1/k$ is a standard
matrix monotone function. In the present notation $\Omega^k_\sigma=(\J_\sigma^{1/k})^{-1}$ and 
for density matrices we have
$$
\chi^2_k(\rho,\sigma)=\< \rho , \Omega^k_\sigma \rho\> -1=
\< \rho , (\J_\sigma^{1/k})^{-1} \rho \> -1=\gamma_\sigma ^{1/k}(\rho,\rho)-1.
$$
Up to the additive constant this is a monotone metric. The monotonicity of the 
$\chi^2$-divergence follows from (\ref{E:Fmon}) and monotonicity is stated as Theorem 4 
in the paper \cite{TRus}, where the important function $k$ is
$$
k_\alpha(x) = \frac{1}{2}\left( x^{-\alpha} + x^{\alpha -1}\right) \quad \mbox{and} \quad
\chi^2_{k_\alpha}=\chi^2_\alpha.
$$
$1/k_\alpha$  is a standard matrix monotone function for $\alpha \in [0,1]$ and $k_\alpha(x)$
is convex in the variable $\alpha$. The latter implies that $ \chi^2_\alpha$ is convex
in  $\alpha$. The $\chi^2$-divergence $ \chi^2_\alpha$ is minimal if $\alpha=1/2$. (It is
interesting that this appeared in  \cite{PD143} as Example 4.)

When $1/k(x)=(1+x)/2$ is the largest standard matrix monotone function, then the 
corresponding $\chi^2$-divergence is the smallest and in the paper  \cite{TRus} 
the notation $\chi^2_{Bures}(\rho,\sigma)$ is used. Actually,
$$
\chi^2_{Bures}(\rho,\sigma)= 2 \int_0^\infty \Tr \rho \exp (-t\omega ) \rho \exp (-t\omega)\,dt
-1 ,
$$ 
see Example 1 in \cite{PD143}.

The monotonicity and the classical inequality (\ref{E:chinorm}) imply
$$
\|\rho-\sigma\|_1^2 \le \chi^2(\rho,\sigma)
$$
(when the conditional expectation onto the commutative algebra generated by $\rho-\sigma$
is used). 


\section{Extension of monotone metric}

As an extension of the operator (\ref{E:jede}),
define $\J_{D_1,D_2}^f :\bM_n \to \bM_n$ as
$$
\J_{D_1,D_2}^f=f(\bL_{D_1}\bR_{D_2}^{-1})\bR_{D_2}\equiv f(\Delta(D_1/D_2))\bR_{D_2}
= \bL_{D_1} \, m_f \, \bR_{D_2} \,,\,,
$$
where $f: \bbbr^+ \to \bbbr^+$. In this terminology
$$
S^A_f (\rho_1\|\rho_2)=\< A, \J_{\rho_1,\rho_2}^f A\>.
$$
Theorem \ref{T:quasiconv} says that for a matrix monotone function $f$, 
$\< A, \J_{\rho_1,\rho_2}^f A\>$ is a jointly concave function of the variables $\rho_1$ 
and $\rho_2$.

The monotone metrics contains $(\J_{\rho,\rho}^f)^{-1}$, therefore we consider 
the inverse
$$
(\J_{D_1,D_2}^f)^{-1}=f^{-1}(\Delta(D_1/D_2))\bR_{D_2}^{-1}.
$$
In this chapter $\beta$ is completely positive trace preserving mapping between matrix spaces.

\begin{lemma}
Assume that $D_1, D_2, \beta (D_1), \beta(D_2)$ are positive definite and $f>0$. 
Then the conditions
\begin{equation}\label{jef1}
\beta^* (\J_{\beta(D_1),\beta(D_2)}^f)^{-1}\beta \le (\J_{D_1,D_2}^f)^{-1}
\end{equation}
and
\begin{equation}\label{jef2}
\beta \J_{D_1,D_2}^f \beta^* \le \J_{\beta(D_1),\beta(D_2)}^f
\end{equation}
are equivalent.
\end{lemma}

\proof
The following inequalities are equivalent forms of (\ref{jef1}):
$$
(\J_{D_1,D_2}^f)^{1/2}\beta^* (\J_{\beta(D_1),\beta(D_2)}^f)^{-1}\beta (\J_{D_1,D_2}^f)^{1/2}\le I
$$ $$
\|(\J_{\beta(D_1),\beta(D_2)}^f)^{-1/2}\beta (\J_{D_1,D_2}^f)^{1/2}\|^2=
\|(\J_{D_1,D_2}^f)^{1/2}\beta^* (\J_{\beta(D_1),\beta(D_2)}^f)^{-1}\beta (\J_{D_1,D_2}^f)^{1/2}\| \le 1 
$$ $$
\| (\J_{D_1,D_2}^f)^{1/2}\beta^* (\J_{\beta(D_1),\beta(D_2)}^f)^{-1/2}\| \le 1
$$ $$
(\J_{\beta(D_1),\beta(D_2)}^f)^{-1/2}\beta (\J_{D_1,D_2}^f)\beta^* (\J_{\beta(D_1),\beta(D_2)}^f)^{-1/2} \le I
$$
The last inequality is equivalent to (\ref{jef2}). \qed

\begin{pl}
Let $f(x)=sx+1$. Then
\begin{eqnarray*}
\< A, (\J_{D_1,D_2}^f)^{-1}A\> &=&\<A, (s\Delta(D_1/D_2)+1)^{-1}\bR_{D_2}^{-1} A\>
=\<A, \left((s\Delta(D_1/D_2)+1)\bR_{D_2}\right)^{-1} A\>\cr
&=&
\<A, (s\bL_{D_1}+\bR_{D_2})^{-1} A\>.
\end{eqnarray*}
This was studied in the paper \cite{L-R}, where the result
\begin{equation}\label{jef21}
\beta^* (s\bL_{\beta(D_1)}+\bR_{\beta(D_2)})^{-1}\beta \le (s\bL_{D_1}+\bR_{D_2})^{-1}
\end{equation}
was obtained. Another formulation is
\begin{equation}
\beta^* (\J_{\beta(D_1),\beta(D_2)}^f)^{-1}\beta \le (\J_{D_1,D_2}^f)^{-1}
\end{equation}
which is equivalent to
\begin{equation}\label{jef22}
\beta \J_{D_1,D_2}^f \beta^* \le \J_{\beta(D_1),\beta(D_2)}^f
\end{equation}
due to the previous Lemma.

For $f(x)=sx+1$ this is rather obvious:
$$
\< A, \beta \J_{D_1,D_2}^f \beta^* A\>=s \Tr D_1 \beta^*(A) \beta^*(A^*)+ \Tr
\Tr D_2 \beta^*(A^*) \beta^*(A)
$$
and
$$
\< A, \J_{\beta(D_1),\beta(D_2)}^f A\>=s \Tr D_1 \beta^*(AA^*)+ \Tr
\Tr D_2 \beta^*(A^*A).
$$
The Schwarz inequality
$$
 \beta^*(X) \beta^*(X^*) \le \beta^*(XX^*)
$$
is needed and gives (\ref{jef21}) and (\ref{jef22}).\qed
\end{pl}

\begin{thm}\label{T:uj}
Let $\beta:\bM_n \to \bM_m$ be a completely positive trace preserving mapping and
$f:[0,+\infty ) \to (0, +\infty)$ be an operator monotone function. Assume that
$D_1,D_2, \beta(D_1),$ $\beta (D_2)$ are positive definite. Then
$$
\beta^* (\J_{\beta(D_1),\beta(D_2)}^f)^{-1}\beta \le (\J_{D_1,D_2}^f)^{-1}.
$$
\end{thm}

\proof
Due to the Lemma it is enough to prove (\ref{jef2}) for an operator monotone function. 
Based on the L\"owner theorem, we consider $f(x)=x/(\lambda+x)$ ($\lambda >0$). So
$$
\J_{D_1,D_2}^f=\frac{\bL_{D_1}}{\lambda I+\bL_{D_1}\bR_{D_2}^{-1}}
$$
and we need (\ref{jef2}). The equivalent form (\ref{jef2}) is
$$
\<\beta (A), (\lambda I+\bL_{\beta(D_1)}\bR_{\beta(D_2)}^{-1})\bL_{\beta(D_1)}^{-1}\beta (A)\> \le
\<A, (\lambda I+\bL_{D_1}\bR_{D_2}^{-1})\bL_{D_1}^{-1}A\>
$$
or
$$
\lambda \Tr \beta(A^*) \beta(D_1)^{-1}\beta (A)+\Tr \beta (A)\beta(D_1)^{-1} \beta(A^*) 
\le \lambda \Tr A^* D_1^{-1} A + \Tr A D_2^{-1}A^*.
$$
This inequality is true due to the matrix inequality
$$
\beta(X^*) \beta(Y)^{-1}\beta (X)\le \beta(X^* Y^{-1}X) \qquad (Y > 0),
$$
see \cite{Lieb-R}. \qed

The generalized monotone metric
\begin{equation}
\gamma^f_{D_1,D_2}(A,B):=\< A, (\J_{D_1,D_2}^f)^{-1}B\>
\end{equation}
is an extension of the monotone metric which is the case $D=D_1=D_2$. We can call it also as
{\it monotone metric with two parameters}. (The use of this quantity is not clear to me 
in the moment, although the case $f(x)=
1+sx$ appeared already in the paper \cite{L-R}.)

\begin{pl}
Let $f(x)=(x+1)/2$. Then
$$
\J_{D_1,D_2}^f B=\fel (D_1B+BD_2)
$$
and
$$
(\J_{D_1,D_2}^f)^{-1}C=\int_0^\infty 
\exp (-tD_1/2) C \exp (-tD_2/2)\,dt.
$$
If $D_1,D_2$ and $C$ commute, then
$$
(\J_{D_1,D_2}^f)^{-1}C=\left(\frac{D_1+D_2}{2}\right)^{-1}C.
$$ \qed
\end{pl}

\begin{pl}
Let $f(x)=(x-1)/\log x$. Then similarly to $\J_{D}^f$, we have
$$
\J_{D_1,D_2}^fA=\int_0^1 D_1^t AD_2^{1-t}\, dt.
$$
When
$$
D_1=\sum_i \lambda_i P_i \quad \mbox{and} \quad D_2=\sum_j \mu_j Q_j
$$
are the spectral decompositions, then
\begin{equation}\label{E:22}
\J_{D_1,D_2}^fA=\sum_{i,j} m_f(\lambda_i, \mu_j) P_iAQ_j,
\end{equation}
where $m_f$ is the logarithmic mean. (The formula is general, it holds for all standard
matrix monotone functions $f$.) To show that
$$
(\J_{D_1,D_2}^f)^{-1}C=\int_0^\infty (D_1+tI)^{-1} C (D_2+tI)^{-1} \,dt.
$$
is really the inverse, we compute
$$
\int_0^\infty (D_1+tI)^{-1} C (D_2+tI)^{-1} \,dt=\sum_{i,j} \frac{1}{m_f(\lambda_i, \mu_j)} P_iCQ_j,
$$

If $D_1,D_2$ and $C$ commute, then
$$
(\J_{D_1,D_2}^f)^{-1}C=\frac{D_1-D_2}{\log D_1-\log D_2}C.
$$ 
We can recognize that in the commuting case
$$
\J_{D_1,D_2}^fC=m_f(D_1,D_2)C, \qquad (\J_{D_1,D_2}^f)^{-1}C=\frac{1}{m_f(D_1,D_2)}C,
$$
where $m_f$ is the mean generated by the function $f$, $m_f(x,y)=x f(y/x)$.
\qed
\end{pl}

\begin{cor}\label{C:uj}
For a matrix monotone function $f$ the generalized monotone metric 
$$
\<A,(\J_{D_1,D_2}^f)^{-1}A\>
$$
is jointly convex function of the variables $D_1$ and $D_2$.
\end{cor}

The difference between two parameters and one parameter is not essential if the 
matrix size can be changed. Let
\begin{equation}\label{E:2}
D=\left[\matrix{ D_2 & 0 \cr 0 & D_1}\right]  , \qquad \mbox{and} \qquad
A=\left[\matrix{ 0 & B \cr B & 0}\right].
\end{equation}
We show that
$$
\< A, \J_{D}^f A\>=\< B, \J_{D_1,D_2}^fB\>+\< B, \J_{D_1,D_2}^g B\>
$$
where $g(x)=xf(x^{-1})$. Since continuous functions can be approximated by polynomials,
it is enough to check $f(x)=x^k$ trivially. The case of inverse functions is similar.

\begin{lemma}
For standard operator monotone function $f$, we have
$$
\< A, \J_{D}^f A\>=2\< B, \J_{D_1,D_2}^fB\> \qquad \mbox{and} \qquad
\< A, (\J_{D}^f)^{-1} A\>=2\< B, (\J_{D_1,D_2}^f)^{-1}B\>
$$
for the matrices (\ref{E:2}).
\end{lemma}

It follows that the monotonicity, Theorem \ref{T:uj}, and the joint convexity, Corollary
\ref{C:uj}, are consequences of  the one parameter case.

\end{document}